\documentclass[twocolumn,aps,showpacs,preprintnumbers,amsmath,amssymb]{revtex4}
\usepackage{graphicx}

\begin{document}

\title{Soliton-sound interactions in quasi-one-dimensional Bose-Einstein condensates}

\author{N.G. Parker, N.P. Proukakis, M. Leadbeater, and C.S. Adams}

\affiliation{Department of Physics, University of Durham, South Road, Durham DH1 3LE, United
Kingdom}
\begin{abstract}

Longitudinal confinement of dark solitons in quasi-one-dimensional Bose-Einstein
condensates leads to sound emission and reabsorption.  We perform quantitative studies of the
dynamics of a soliton oscillating in a tight dimple trap, embedded in a weaker
harmonic trap.  The dimple depth provides a sensitive handle to control the soliton-sound
interaction. In the limit of no reabsorption, the power radiated is found to be proportional to the
soliton acceleration squared.  An experiment is proposed to detect sound emission as a change in
amplitude and frequency of soliton oscillations.

\end{abstract}
\pacs{03.75.Lm, 42.65.Tg}
\maketitle

The experimental realisation of dilute atomic Bose-Einstein condensates (BEC) \cite{Fermi}
has introduced an unparalleled platform from which to study the dynamical behaviour of
nonlinear systems. Of great interest is the stability of topological structures,
such as vortices and dark solitary waves. Recent experiments at low temperatures
have highlighted the temperature-independent crystallization \cite{Abo} and decay \cite{Davies} of vortex
structures.  Related theoretical work has indicated the importance of vortex-sound interactions 
\cite{Vinen}. Following the recent
experimental observation of dark solitary waves (henceforth referred to as dark solitons)
in BEC's \cite{Burger}, one can now study the coupling of solitons to the 
background sound field.

Dark solitons are robust localised one-dimensional defects characterised by a notch in the 
ambient condensate density and a phase slip across the centre.  They are supported in
repulsive (defocussing) nonlinear media where the kinetic dispersion of the wave is balanced
by the nonlinear interaction. In three-dimensional geometries they are prone to decay into
lower energy topological structures such as vortex rings \cite{AndersonB}.
However, in quasi-one-dimensional (quasi-1D) geometries, where the transverse condensate size
is of the order of the healing length, solitons are expected to exhibit longer lifetimes, due
to the suppression of transverse excitations \cite{Muryshev1}.

In the context of nonlinear optics \cite{Kivshar}, dark solitons on a homogeneous
background can become intrinsically unstable to modifications of
the nonlinearity within the optical medium (e.g. due to saturation effects)
\cite{Pelinovsky}. In quasi-1D BEC's, modified nonlinearities can arise
from the transverse dimensions that have been integrated out, as considered for
longitudinally homogeneous condensates in \cite{Muryshev2}. 
The longitudinal confinement of current atomic BEC experiments leads to an additional 
mechanism by which the
integrability of the system can be broken \cite{Busch,Fedichev,Huang}. In this paper we
 perform detailed quantitative analysis of 
the stability of dark solitons in quasi-1D geometries under longitudinal confinement,
and find that dissipation due to the longitudinal inhomogeneity dominates over decay arising
from coupling to transverse modes \cite{Muryshev2}. Our axial configuration consists of a
tight dimple trap (where the soliton resides) embedded in a weaker harmonic potential. Such a
geometry enables us to control and examine the interplay between 
sound emission and reabsorption by the oscillating
soliton. In the limit of no reabsorption, the rate of emission is found to be
proportional to the local soliton acceleration squared, as predicted for optical solitons
in {\em homogeneous} nonlinear waveguides \cite{Pelinovsky}.  Finally we
propose a controllable experiment where the soliton-sound interactions can be quantified by measuring
the amplitude and frequency of the soliton oscillations.

Our analysis is based on numerical simulations of the Gross-Pitaevskii Equation (GPE), describing
the dynamics of weakly-interacting BEC's in the limit of low temperature,
\begin{eqnarray}
i\hbar \frac{\partial \psi}{\partial t}=-\frac{\hbar^2}{2m}\nabla^2
\psi+V_{\rm{ext}}\psi+g|\psi|^2 \psi.
\end{eqnarray}
Here $\psi$ is the order parameter of the system, $V_{\rm{ext}}$ the confining
potential, $m$ the atomic mass and $g=4\pi\hbar^2a/m$ the scattering amplitude, where $a$ is the
{\it s}-wave scattering length.  In the limit of tight transverse confinement, the condensate
dynamics are effectively longitudinal \cite{Jackson}.  By integrating out the transverse
degrees of freedom, the GPE reduces to a one-dimensional ($1$D) form with a modified coupling
coefficient
$g_{_{\rm 1D}}=g/(2\pi l_{_\perp}^2)$, where $l_{_\perp}=\sqrt{\hbar/m\omega_{_\perp}}$ is
the transverse harmonic oscillator length and $\omega_{_\perp}$ the
transverse trapping frequency.  On a homogeneous background density $n_{_{\rm 1D}}$, a dark soliton of
speed $v$ and position, $z'=(z-vt)$, is defined as,
\begin{eqnarray}
\psi(z,t)=\sqrt{n_{_{\rm 1D}}}
\left[\beta \tanh
(\beta z'/\xi)+i
(v/c)
\right]{\rm e}^{-i\mu_{_{\rm 1D}}t/\hbar}
\end{eqnarray}
where $\mu_{_{\rm 1D}}=g_{_{\rm 1D}}n_{_{\rm 1D}}$ is the
one-dimensional chemical potential,
$c=\sqrt{\mu_{_{\rm
1D}}/m}$ the Bogoliubov speed of sound, $\beta=\sqrt{1-(v/c)^2}$, and $\xi=\hbar/\sqrt{m 
\mu_{_{\rm 1D}}}$ is the
healing length characterising the size of the soliton.

We consider a trap of the form
\begin{equation}
V_{\rm ext}=\left\{
\begin{array}{c}
\frac{1}{2}m\omega_{z}^2z^2 \\
 V_0+\frac{1}{2}m\omega_{\zeta}^2(|z|-z_0)^2
\end{array}
\right.
{\rm for}
\begin{array}{cr}
& |z|\leq z_0 \\
& |z|> z_0
\end{array}
\end{equation}
where we have defined the `cut-off' depth of the inner `dimple' trap as 
$V_0=\frac{1}{2}m\omega_{z}^2z_0^2$ (see Fig.~1).
Sound waves generated by the oscillating soliton can either be trapped within the dimple
or escape to the outer trap, depending on the value of the cut-off $V_0$ relative to the
chemical potential $\mu_{_{\rm 1D}}$. For a sufficiently weak outer trap, the density in the outer
region and close to
the dimple becomes approximately homogeneous (Fig.~1(b)), and as a first
approximation we initially solve the 1D GPE in the limit $\omega_{\zeta}=0$. We shall
return to the experimentally relevant case of  $\omega_\zeta \neq 0$ and 3D simulations
later. The effect of the emitted sound on soliton dynamics can be determined by
monitoring the `soliton energy' \cite{Soliton_Energy}, 
a quantity  determined by integrating the functional,
\begin{eqnarray}
\varepsilon(\psi)=\frac{\hbar^2}{2m}\left|\nabla\psi\right|^2+V_{\rm
ext}\left|\psi\right|^2+\frac{g}{2}\left|\psi\right|^4,
\end{eqnarray}   
across the soliton region ($z_{\rm s}\pm 5\xi$) and subtracting the corresponding
contribution of the background fluid.

\begin{figure}
\includegraphics[width=8.2cm]{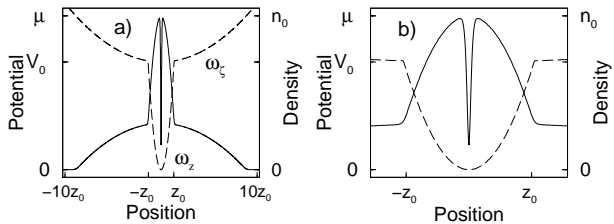}
\caption{ (a) Initial longitudinal condensate density (solid line) for a double trap geometry
(dashed line)
with a
soliton centred at the origin. (b) Enlarged image of central region.}
\end{figure}

Fig.~2 shows the temporal evolution of the soliton energy for different cut-offs
(corresponding to different amounts of sound reabsorption), all of which
display  a periodicity of approximately 
 $\omega_{z}/\sqrt{2}$ due to the motion of the soliton in the trap
\cite{Busch,Fedichev,Huang}. 
One can discriminate two limiting cases: 
(i) For sufficiently high cut-offs (here $V_0\geq1.2\mu_{_{\rm 1D}}$)
all emitted sound becomes trapped in the dimple, leading principally to the excitation of the dipole
mode \cite{Busch}.
The combined effect of the
dipole mode and soliton, oscillating at different frequencies, induces beating 
of characteristic period
$\tau\approx22$ (see Fig.~2). 
(ii) For low cut-offs ($V_0 \leq 0.4\mu_{_{\rm 1D}}$) there is essentially pure emission (no
reabsorption) of sound by the soliton.
This behaviour is consistent with the change in energy
experienced by a classical particle of negative mass oscillating in a harmonic trap,
subject to a dissipative force which induces an energy loss proportional to the
acceleration squared \cite{Next_Paper}. 
The observation that the soliton dynamics changes significantly around 
$V_0\approx\mu_{_{\rm 1D}}$ (since sound excitations have energy of
the order of the chemical potential) provides a sensitive handle on the
soliton-sound interaction, which can be controlled experimentally.

\begin{figure}
\includegraphics[width=8.cm]{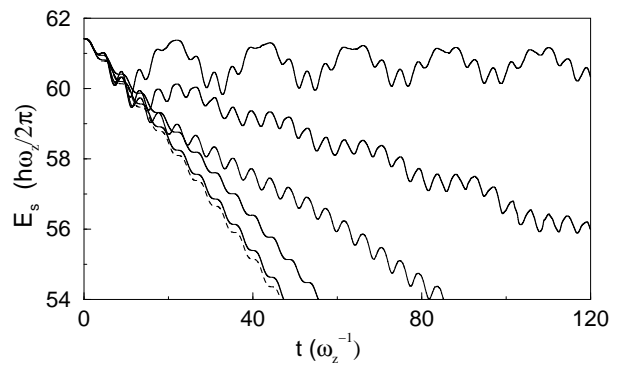}
\caption{Soliton energy versus time for a soliton of initial speed $0.5c$
and position $z=0$ in a relatively weak inner harmonic trap ($\mu_{_{\rm
1D}}=70\hbar\omega_{z}$) for various potential cutoffs $V_0$ (from top to bottom):
$1.1\mu_{_{\rm 1D}}$, $1.02\mu_{_{\rm 1D}}$, $1\mu_{_{\rm 1D}}$, $0.8\mu_{_{\rm 1D}}$
and $0.4\mu_{_{\rm 1D}}$. The lowermost dotted line corresponds to the theoretical
prediction of Eq.~(5), valid in the limit of no reabsorption.}
\end{figure}

The dissipative dynamics of a dark soliton in a shallow dimple trap are
investigated in Fig.~3. 
As the soliton oscillates, it becomes asymmetrically deformed \cite{Next_Paper}
and tries to adjust to the
inhomogeneous background by radiating counter-propagating sound pulses (Fig.~3(a)).
Fig.~3(b) shows the power emitted by the soliton as a function of time. At the bottom of
the trap there is no sound emission since Eq.~(2) becomes the local solution to the
GPE, while at the zenith of the first few oscillations, the soliton radiates maximum
power. This sound emission (which modifies the soliton depth)
leads to an increase in the soliton speed and amplitude of
oscillations, which in turn causes a gradual increase in the peak power radiated.  As
the soliton approaches the edge of the dimple, it experiences
a smoothing of the effective potential due to fluid healing. This reduces the
emission at the trap edge, as manifested in the `intermediate' dips at $\omega_{z} t
\approx 25$ and $29.5$.  Eventually at $\omega_{z} t \approx 35$ the soliton escapes the
trap \cite{Busch}.

In the context of nonlinear optics, multiscale asymptotic techniques 
predict that an unstable optical soliton, propagating in a homogeneous 
nonlinear waveguide, emits sound at a rate proportional to the soliton acceleration squared \cite{Pelinovsky}, such that,
\begin{eqnarray}
\frac{dE_{\rm s}}{dt}&=&-\frac{c}{c^2-v^2}\left[\frac{2c^2}{n}\left(\frac{\partial
N_{\rm s}}{\partial
v}\right)^2\right.
\\   
\nonumber
&~&\left.+2v\left(\frac{\partial
N_{\rm s}}{\partial v}\right) \left(\frac{\partial S_{\rm s}}{\partial
v}\right)+
\frac{n}{2}\left(\frac{\partial S_{\rm 
s}}{\partial
v}\right)^2 \right]\left(\frac{dv}{dt}\right)^2.
\end{eqnarray}
Here $n$ is the local background density, $S_{\rm s}$ the total phase slip  across the
moving soliton, and $N_{\rm s}=\int\left(n-|\psi|^2\right){\rm{d}}z$ the number of
particles displaced by the soliton. In the optical case, the breakdown of integrability
leading to sound emission arises from modifications in the nonlinearity within the
medium, as opposed to the longitudinal confinement discussed above. 
\begin{figure}
\includegraphics[width=7.9cm]{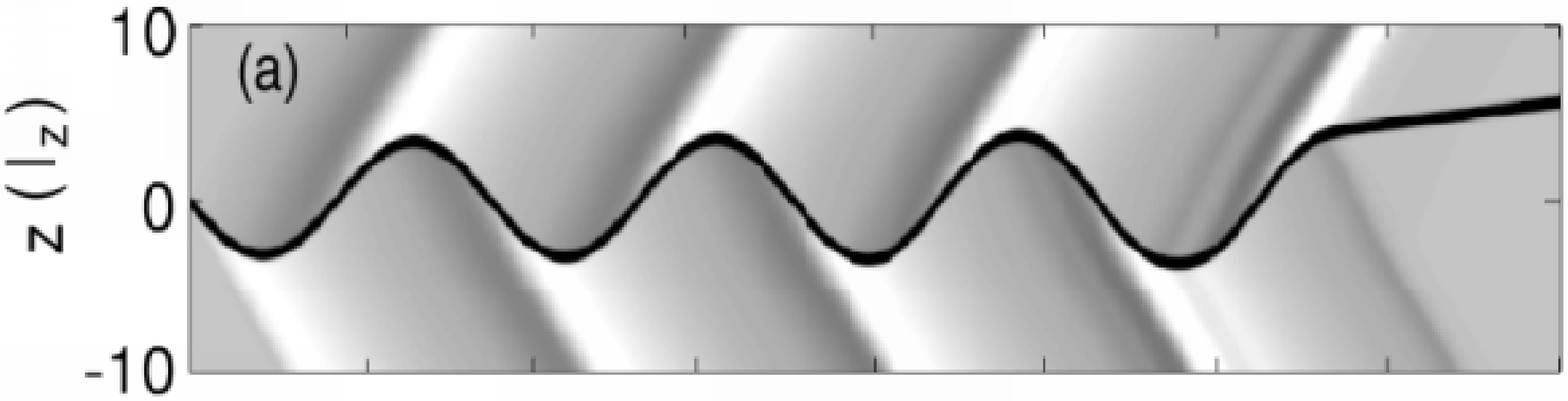}
\\
\includegraphics[width=8.cm]{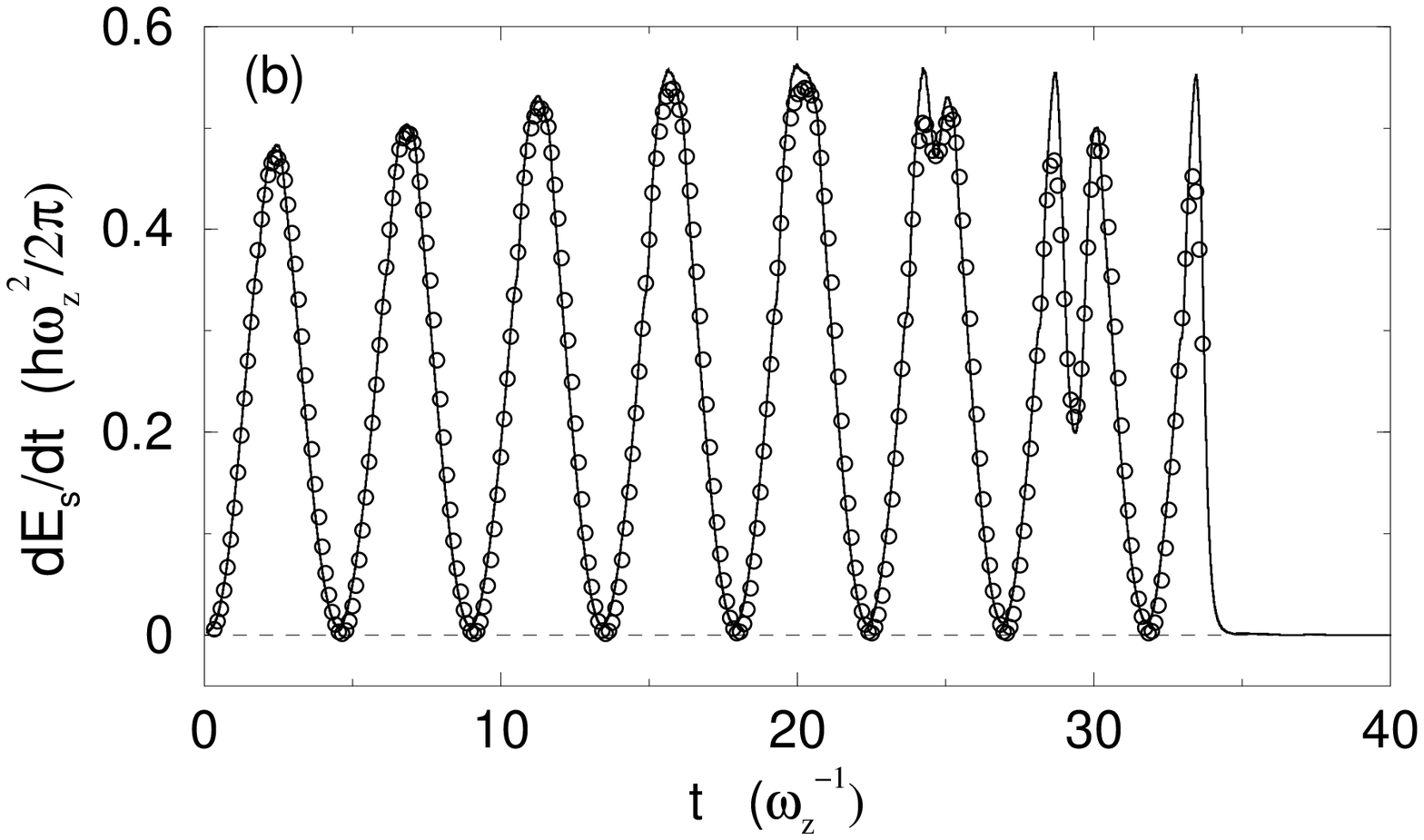}
\caption{ (a) Space-time plot of a soliton oscillating in a dimple trap with
$V_0=0.4\mu_{_{\rm 1D}}$ where  $\mu_{_{\rm 1D}}=35\hbar\omega_{z}$. Deviations from
background density correspond to sound (light/dark regions indicate  high/low
densities), with the sound amplitude roughly $2\%$ of the background.  (b) Power emitted
by the soliton as calculated from the energy functional (solid line), compared
with the prediction of Eq.~(5) (circles).}
\end{figure}
It may therefore come as a surprise that this expression could also be applicable to the
inhomogeneous case. In fact, this can only be valid in double trap structures of Eq.~(3)
if the dimple trap is sufficiently shallow that it essentially allows
complete escape of sound to the outer trap; even then, this result will only
hold until the emitted sound returns to the dimple region
after reflection from the outer trap. In this limit,
the asymptotic conditions become practically indistinguishable from the homogeneous
ones, thus justifying the above approach.
The validity of Eq.~(5) is not dependent on the origin of the instability, but rather on the fact that the
emitted sound escapes from the soliton region.  Note that the effect of the dimple confinement
is implicit in the variations of the parameters $S_{\rm s}$, $c$ and $n$, and, contrary
to the homogeneous case \cite{Pelinovsky}, this result will hold for {\it all} soliton speeds,
due to the continuously-induced `soliton instability' caused by the longitudinal
confinement. Fig.~3(b) confirms the applicability of Eqs.~(5) (circles) in
the limit $V_0 \ll \mu_{_{\rm 1D}}$ by direct comparison to  the 1D GPE
simulations (solid line). Further evidence is provided in Fig.~2 which shows clearly
that as the cut-off is decreased, the variation of energy with time (solid lines)
approaches the above  analytical result in the limit of no reabsorption (dotted line).
We stress that Eq.~(5) will {\it not} hold in the usual infinite
harmonic traps, where multiscale asymptotic technique must be combined with
boundary layer theory \cite{Busch}.

\begin{figure}
\includegraphics[width=8.2cm]{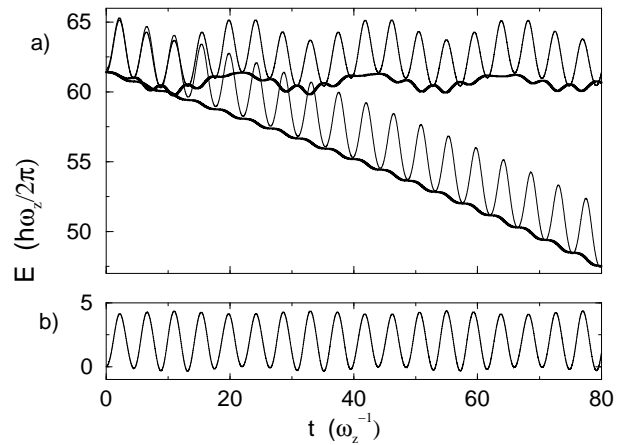}
\caption{(a) Soliton energies obtained from the 3D GPE (thin lines) vs.~corresponding 1D
GPE results (thick lines) for cut-offs of $V_0=1.1\mu_{_{\rm 1D}}$ (top) and
$V_0=0.4\mu_{_{\rm 1D}}$ (bottom). 
The longitudinal parameters are as in Fig.~2, and trap aspect ratio is set to
$\omega_\perp/\omega_{z}\approx250$, such that the initial density profile in the {\it
z}-direction is identical to the one-dimensional density.
(b) The oscillation in the total longitudinal energy due to the coupling to transverse
modes, as measured by integrating the energy functional longitudinally. }
\end{figure}

Simulations of the 3D GPE
with cylindrical symmetry (still under the assumption of a homogeneous outer trap)
confirm the qualitative findings of Fig.~2. In particular, in the limit 
$\mu_{_{\rm 3D}} / \hbar \omega_{_{\perp}} \approx 1.2$ (where
the soliton was recently
predicted to be stable against transverse decay \cite{Muryshev2})
we find the dominant decay mechanism to be due to {\it axial} inhomogeneity.
The coupling to transverse modes \cite{Huang} merely leads to an additional {\it oscillation}
of the soliton energy in time (Fig.~4), whose amplitude increases as the transverse
confinement is relaxed \cite{Comment3}. 
Subtracting the average oscillation in the 3D
longitudinal energy (Fig.~4(b)) from the energy of the soliton region gives
essentially perfect agreement with the 1D results (Fig.~2).

Finally we extend our earlier treatment to the experimentally relevant case of
$\omega_\zeta\neq0$. We consider a BEC in a quasi-1D 
geometry \cite{Gorlitz} in a
cylindrically symmetric trap with frequencies $\omega_\zeta$ (longitudinal) and
$\omega_\perp$ (transverse), and then add an inner dimple potential ($\omega_z$), such that the
longitudinal confinement is as shown in Fig.~1. This could be realized, for example, by
a dimple microtrap created optically within a larger magnetic trap. The key points
for such a realization are: (i) a very weak longitudinal trap frequency $\omega_\zeta$,
such that the sound escaping from the inner trap is not reflected back from the outer
trap within the timescale of interest, and (ii) a sufficiently tight inner trap
$\omega_{z}$ such that the oscillating soliton emits sound at a rate fast enough that
its amplitude increases noticeably before other decay mechanisms become important.

\begin{figure}
\includegraphics[width=8.cm]{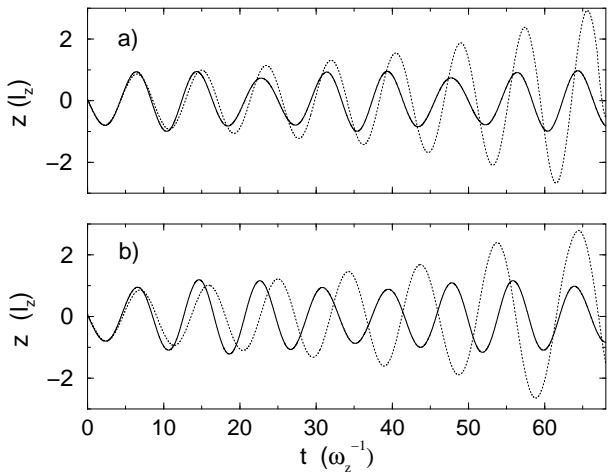}
\caption{Soliton path in a longitudinal double-trap configuration for (a) tight harmonic
inner trap and (b) gaussian dimple $V_0 [1-\exp(- \alpha z^2)]$. Each graph shows the
cases of a cut-off $V_0 =\mu_{_{\rm 1D}}$ (dotted line) compared to an effectively
infinite cut-off $V_0 =5 \mu_{_{\rm 1D}}$ (solid line), for a soliton with initial speed
$0.2 c$. The gaussian dimple in (b) is chosen such that it yields same initial
oscillation frequencies as the harmonic dimple (achieved for $\alpha = 
\omega_z^2/2V_0
$). The outer trap is harmonic in both cases, with relatively weak confinement
$\omega_\zeta=\omega_z/10$. }
\end{figure}

Experimentally, the effect of soliton-sound interactions can be probed by determining
the soliton path via repeated time of flight measurements, and comparing paths for
different depths of the dimple trap. The best signature for this effect arises when 
comparing the case of an effectively
infinite cut-off to that of a lower cut-off, which nonetheless permits a sufficient
number of soliton oscillations. Since the soliton rapidly escapes the
trap for $V_0 < \mu_{_{\rm1D}}$, we thus propose the use of  $V_0 = \mu_{_{\rm1D}}$ as a
low cut-off, for which the soliton remains confined and ultimately decays to a sound
pulse. Such a comparison has been performed for the double-trap system in Fig.~5, for
both harmonic and gaussian dimple traps. One clearly observes a change in both amplitude
and frequency of the soliton oscillations as a result of sound emission. These effects
become more noticeable in the gaussian trap, for which the required cut-off for total
sound entrapment is larger than the harmonic one. Choosing `typical' parameters
$\omega_\zeta=2 \pi \times 5$ ${\rm Hz}$, $\omega_{z} = 10 \omega_{\zeta}$ and
$\omega_{_{\perp}} = 250 \omega_{\zeta}$ and $\mu_{_{\rm 3D}} = 8 \hbar 
\overline{\omega}$ where $\overline{\omega} = (\omega_{\zeta} 
\omega_{_{\perp}}^{2})^{1/3}$, we find respectively for a $^{23}{\rm Na}$ ($^{87}{\rm
Rb}$) condensate, a one-dimensional peak dimple density of $n_{_{\rm 1D}} = 5 \times
10^{7}$ ($ 1.5 \times 10^{7}  $) ${\rm m}^{-1}$, condensate atom number $N= 18,000$
(3,500) and a total observation timescale for Fig.~5 of $\tau_{\rm exp}=220$
${\rm ms}$, consistent with the expected soliton lifetime in such systems
\cite{Muryshev2,Busch,Fedichev}.

 In summary, we have performed quantitative studies of 
the dynamics of dark solitons oscillating in trapped quasi-one-dimensional Bose condensates
under longitudinal double-trap confinement. This geometry is ideal for controlling the interplay between
sound emission and reabsorption, and we have identified two limiting cases: (i) In the limit of
no reabsorption, the
power emitted by the soliton is proportional to the square of its  acceleration.
(ii) The opposite limit of
effectively infinite traps (such as harmonic traps in which soliton experiments have
been performed) leads to the anticipated stabilization against soliton decay.
 We suggest that the sound
radiation can be quantified by tracking the position of the dark soliton in
suitably engineered traps, which allow for precise control of the soliton-sound interaction.

We acknowledge discussions with C. Barenghi, K. Burnett, W.D. Phillips and G.V. 
Shlyapnikov. Funding was provided by UK EPSRC.


\begin{thebibliography}{99}
\bibitem{Soliton_Energy} This procedure cannot discriminate between soliton and
sound energy present in the interval; however, this sound contribution is
much smaller than the soliton energy.
\end{thebibliography}
\end{document}